\documentclass[pra,reprint,superscriptaddress,showkeys]{revtex4-2}
 
\usepackage[letterpaper, margin=.75in]{geometry}
\usepackage{graphicx} 
\usepackage{multirow}%
\usepackage{amsmath,amssymb,amsfonts}%
\usepackage{amsthm}%
\usepackage{mathrsfs}%
\usepackage[title]{appendix}%
\usepackage{xcolor}%
\usepackage{textcomp}%
\usepackage{booktabs}%
\usepackage{algorithm}%
\usepackage{algorithmicx}%
\usepackage{algpseudocode}%
\usepackage{listings}%
\usepackage{epstopdf}
\usepackage{tabularx}
\usepackage[pdfpagelabels=true]{hyperref}
\usepackage{hypernat}
\usepackage[nolist]{acronym}
\usepackage{fancyhdr}
\usepackage{shortcuts} 
\usepackage{xr}
\usepackage{blindtext}

\def\ttitle{From MOT to BEC using a single crossed-wire pair}
\def\kkeywords{
  cold atoms, quantum sensing, magneto-optical traps
}

\setboolean{showcomment}{true}

\usepackage{lineno}
\hypersetup{
    naturalnames=true,
    colorlinks=true,
    linkcolor=blue,
    pdfpagemode=UseNone,
    pdfstartview=FitH,
    pdftitle={\ttitle},
    pdfauthor=Air Force Research Laboratory,
    pdfsubject={Machine learning for experimental atomic physics},
    pdfkeywords={\kkeywords},
}

\newcolumntype{L}[1]{>{\raggedright\let\newline\\\arraybackslash}m{#1}}
\newcolumntype{C}[1]{>{\centering\let\newline\\\arraybackslash}m{#1}}
\newcolumntype{R}[1]{>{\raggedleft\let\newline\\\arraybackslash}m{#1}}

\newcommand{\distA}[1]{%
  Approved for public release; distribution is unlimited.  Public Affairs %
  release approval %
  #1.
}

\begin{document}

\title{\ttitle}

\author{Joshua M Wilson}
\author{James Stickney}
\author{Francisco Fonta}
\author{Johnathan White}
\affiliation{Space Dynamics Laboratory, Quantum Sensing \& Timing, North Logan, UT 84341, USA}

\author{Brian Kasch}
\author{Spencer E. \surname{Olson}}
\affiliation{\AFRLAddress}
\author{Matthew B. \surname{Squires}}
\def\AFRLAddress{%
  Space Vehicles Directorate, %
  Air Force Research Laboratory, %
  3550 Aberdeen Ave SE, %
  Kirtland Air Force Base, %
  87117, %
  New Mexico, %
  USA%
}
\affiliation{\AFRLAddress}
\date{\today}

\begin{abstract}
We demonstrate a new \ac{MOT} configuration using a simple pair of crossed wires
rotated at 45$^\circ$ and an appropriate bias field to generate a \ac{MOT} of
$\mathord{\geq} 10^8$ atoms. The same pair of wires with slightly adjusted control
parameters is then used to magnetically trap the atoms and then cool them with
forced evaporative cooling into a \ac{BEC} with $\mathord{\geq}10^4$ atoms. The theory
for generating a quadrupole field with a pair of crossed wires with an arbitrary
rotation angle is presented. Finally, we present the atom chip design and
fabrication for the crossed-wire pair, as well as the experimental operating
protocols for BEC production using only a single crossewire atom chip.
\end{abstract}

\keywords{\kkeywords}

\maketitle \thispagestyle{fancy}

\begin{acronym}
\acro{2D}{two-dimensional}
\acro{2D+ MOT}{two-dimensional plus MOT}
\acro{3D}{three-dimensional}
\acro{AlN}{aluminum nitride}
\acro{BEC}{Bose-Einstein condensate}
\acro{Cu}{copper}
\acro{DBC}{direct-bonded copper}
\acro{MOT}{magneto-optical trap}
\acro{PGC}{polarization-gradient cooling}
\end{acronym}

\acresetall
\section{Introduction}

Ultra-cold atoms have become essential tools in the development of quantum
technologies. When cooled to temperatures near absolute zero, these atoms
exhibit quantum behaviors that have been harnessed in quantum computing
\cite{wintersperger2023_neutralQC,evered2023_highFidelityGates}, simulation
\cite{halimeh2025_CAQuantSimGaugeTheory,schafer2020_toolsForQS}, sensing
\cite{cohen2022_CAMagnetometer,geiger2020_InertialMeasureCA}, navigation
\cite{abend2023_coldAtomRoadmap,EU2020_CAIs}, and precision measurements
\cite{zhang2016_precisionMeasurement}. The ability to generate ultra-cold atoms
is crucial for advancing these technologies.

Atom chips are one option for producing ultra-cold atoms, provide a compact
and scalable platform, and have become the platform for many quantum sensing
technologies\cite{Keil10102016_atomChipReview}. They have been utilized in
experiments ranging from clocks~\cite{szmuk2015_StabilityChipClock} to
gravimeters \cite{li2023_atomChipGravimeters} and sub-orbital ultra-cold atom
experiments \cite{Lachmann2021_BEC_soundingRocket}. The first step in generating
ultra-cold atoms is pre-cooling in a \ac{MOT} to the order of a few $\mu$K. The
\ac{MOT} is typically made within a few mm of the atom chip to improve
mode matching into the magnetic trap. This may be accomplished using a
mirror-MOT configuration with a quadrupole field produced by a
U-wire~\cite{PhysRevLett.83.3398_mirrorMOT}.

In previous work, we used multiple types of atom chips to generate the various
magnetic fields for different parts of the experimental sequence: a \ac{DBC}
atom chip with a U-wire to produce a quadrupole field for the \ac{MOT} stage and
a second \ac{DBC} atom chip with crossed wires for an Ioffe-Prichard-type magnetic
trap \cite{squires2011_atomChipsDBC, squires2016_exVacuuo}. In the process of
testing this atom chip design, we accidentally discovered that a \ac{MOT} could be
produced with the crosswire magnetic trap wires when the U-wire was turned off.
This was unexpected since the crosswire trap was typically configured to produce
an Ioffe-Prichard-type trap with a non-zero bottom field that was rotated
45$^{\circ}$ relative to the laser cooling beams~\cite{squiresDissertation2008,
PhysRevA.107.063305,farkas2013efficient,horikoshi2006atom}. 

In further testing, we found this \ac{MOT} was as robust as \acp{MOT} made with
coils or U-wires, and we now call this new configuration a crosswire-\ac{MOT}.
We have since empirically optimized the crosswire-MOT, worked out the
theoretical framework explaining the crosswire-MOT, and rebuilt the atom chip
for the experiment without the U-wire. This new atom chip configuration reduces
the number of atom chips that need to be fabricated, improves the net thermal
conductivity of the atom chip to the heat sink, and improves the overlap of the
\ac{MOT} with the magnetic trap.

In this paper, we present a simple design for a crossed-wire atom chip, capable
of creating both a mirror-MOT and a magnetic trap using only a single pair of
crossed wires (in conjunction with standard bias-field coils). We first discuss
a theoretical framework for making a quadrupole \ac{MOT} field with an
appropriate orientation relative to the atom chip. Next, we discuss the
experimental apparatus used for implementing the crosswire-MOT, including a
brief discussion of our current techniques in the use of \ac{DBC} to create an
ex-vacuuo atom chip. Then, we present details in the cooling and trapping
protocols implemented to create a \ac{MOT} of $\mathord{\geq}10^8$ atoms. Finally, using
the same wires, we demonstrate loading into a magnetic trap and evaporation,
achieving a \ac{BEC} with $\mathord{\geq}10^4$ atoms. 

\begin{figure}[h!] \centering
    \includegraphics[width=1\linewidth]{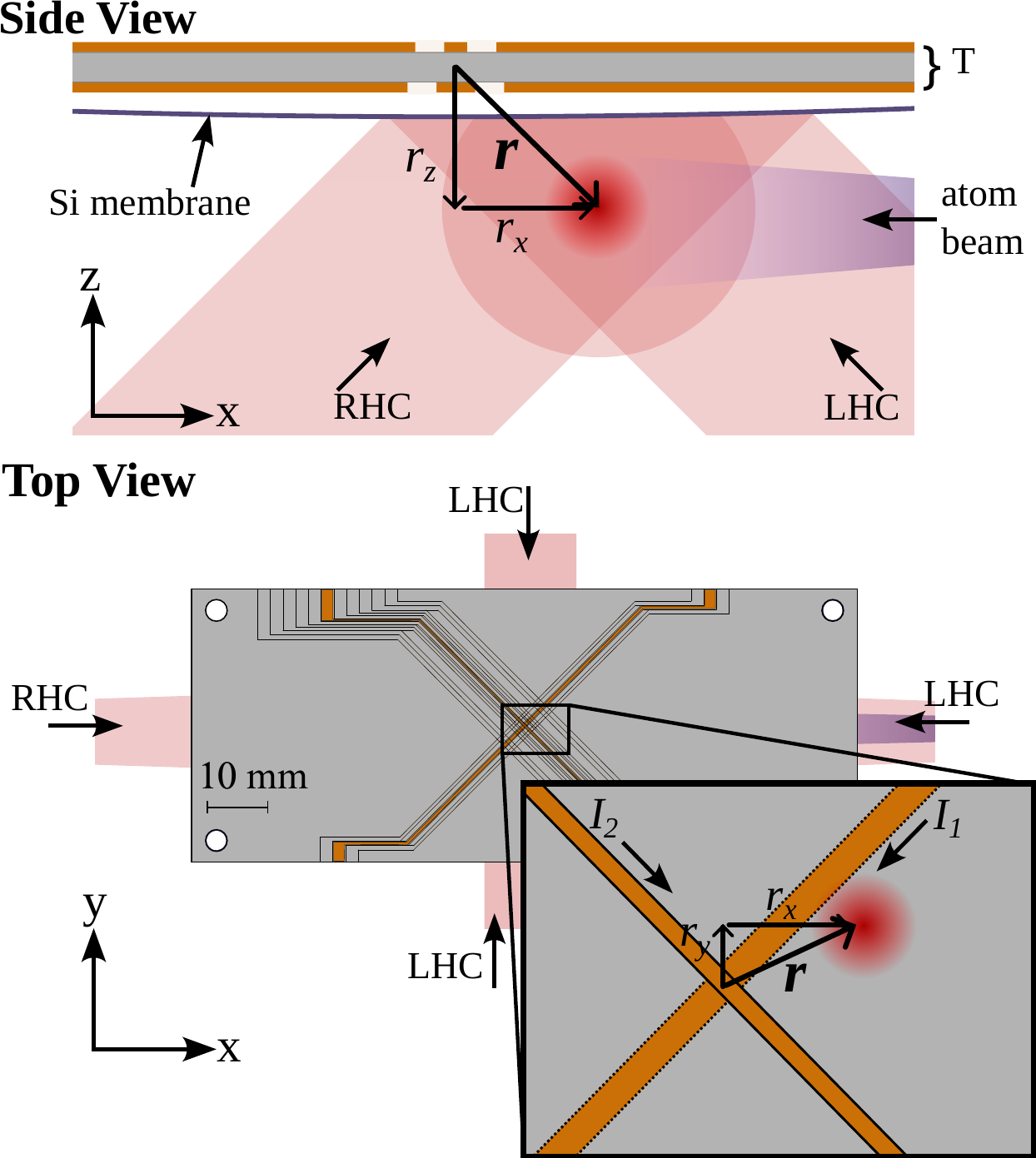}
    \caption{ Representation of the side and top view of the atom chip
    experiment showing the relative locations of the \ac{DBC} atom chip, silicon
    membrane, current-carrying wires, lasers, the atomic flux from a
    2D~\ac{MOT} that is not shown, and the 3D~\ac{MOT}. The representation of
    the side view cross section of the wires on the atom chip is schematic and
    is intended to show the offset of the \ac{MOT} from the intersection of the
    wires. The red cloud represents the approximate location of the \ac{MOT}
    relative to the chip. The indicators LHC and RHC represent left-handed and
    right-handed circular polarization of the laser beams, respectively. In the
    side view, note the curvature of the membrane due to atmospheric pressure on
    the outside. The relative thickness of the membrane and the \ac{DBC} is to
    scale. In the top view, the atom chip is shown with all of the traces.
    Only the wires used in the paper are colored; the other wires are shown only
    in outline. In the zoomed in view, only the wires used for making a \ac{MOT} are
    shown. The dashed outline represents the trace that is on the bottom layer
    of \ac{Cu}, while the solid outline represents the trace on the top layer.
    The red cloud represents the approximate $x,~y$ location of the \ac{MOT}
    relative to the crossed wires. The vector $\mathbf{r}$ starts in the middle
    of the \ac{DBC} stack, and the center of the cross shows the offset of
    the quadrupole zero from the crossing of the wires.}
    \label{fig:SetupCartoon} \end{figure}

\section{Theory of Operation}\label{sec:Theory}

This section develops a linearized description of the magnetic field near the
MOT center generated by two straight, orthogonal conductors patterned on
opposite faces of the \ac{DBC} chip. Our goal is to show how the superposed
fields, together with a uniform bias field, can realize a mirror–MOT with a
prescribed axial gradient and a tunable in-plane asymmetry. We work in the
regime where the cloud samples a region small compared with its distance to the
chip, so the field is well approximated by its value and gradient at the MOT
center. We express the resulting gradient tensor in terms of wire currents and
placement and identify simple current–geometry conditions that reproduce a
rotated anti-Helmholtz form.

The atom chip is an aluminum nitride (AlN) substrate of thickness $T$ with
straight, laser-cut conductors on the two opposite faces. We take
$\hat{\pmb{x}}$ and $\hat{\pmb{y}}$ to lie in the chip plane and $\hat{\pmb{z}}$
to point from the lower to the upper face. The wire carrying current $I_1$ lies
on the $+z$ face ($z=+T/2$) and runs along the in-plane unit vector
$\hat{\pmb{u}}_1=-(\hat{\pmb{x}}+\hat{\pmb{y}})/\sqrt{2}$. The wire carrying
$I_2$ lies on the $-z$ face ($z=-T/2$) and runs along
$\hat{\pmb{u}}_2=(\hat{\pmb{x}}-\hat{\pmb{y}})/\sqrt{2}$. We place the origin at
the chip mid-plane, so the two straight current lines pass through $(0,0,+T/2)$
and $(0,0,-T/2)$, respectively. Figure \ref{fig:SetupCartoon} shows the location of
the \ac{MOT} relative to our atom chip, the laser beams, and our chosen coordinate
axes.

Since we are mostly concerned with the situation where the length scale over
which the atomic cloud samples the magnetic field is much smaller than the
distance between the wires and the cloud, $\pmb{r} \gg \delta\pmb{r}$, the field
can be approximated to first order as
\begin{equation}
  \pmb{B}(\pmb{r}+\delta\pmb{r})
  = \pmb{B}(\pmb{r}) + \hat{\pmb{G}}\cdot\delta\pmb{r},
  \label{eq:linearize}
\end{equation}
where $\hat{\pmb{G}}$ is a traceless $3\times 3$ gradient tensor with components
$G_{ij} = \partial B_i / \partial x_j$. If the principal axes of the gradient
matrix are aligned with the coordinate axes, then $\hat{\pmb{G}}$ is diagonal.

For later comparison to our crosswire-MOT field, a traditional \ac{MOT} field created
by an anti-Helmholtz coil pair is
\begin{equation}
  \hat{\pmb{G}} =
  G\begin{pmatrix}
    1 + \alpha & 0 & 0 \\
    0 & 1 - \alpha & 0 \\
    0 & 0 & -2
  \end{pmatrix},
  \label{eq:AH}
\end{equation}
where $G$ is the axial gradient (set by coil geometry and current), and $\alpha$
is a dimensionless asymmetry parameter. For a first-order optimized Helmholtz
coil pair, $\alpha=0$. However, we explicitly define $\alpha$ to be variable
since the $45^\circ$ crosswire geometry will ultimately necessitate $\alpha\neq
0$.

For a mirror-MOT configuration, the field Eq.~\ref{eq:AH} must be rotated about
the $y$ axis by an angle $\theta$:
\begin{equation}
  \hat {\pmb G}=
  G\begin{pmatrix}
    \tfrac{\alpha - 1}{2} + \tfrac{\alpha + 3}{2} \cos 2\theta & 0 &
    \tfrac{\alpha + 3}{2} \sin 2\theta \\
    0 & 1 - \alpha & 0 \\
    \tfrac{\alpha + 3}{2} \sin 2\theta & 0 &
    \tfrac{\alpha - 1}{2} - \tfrac{\alpha + 3}{2} \cos 2\theta
  \end{pmatrix},
  \label{eq:rotated}
\end{equation}
where we used the tensor transformation rule
$\hat{\pmb{G}}\;\to\;R_y^\dagger(\theta)\,\hat{\pmb{G}}\,R_y(\theta)$. While an
ideal mirror-MOT has $\theta = 45^\circ$, we have left both $\alpha$ and
$\theta$ variable in Eq.~\ref{eq:rotated} since a crosswire-MOT cannot create
the ideal gradient of a mirror-MOT. However, we will show below that a
compromise between $\alpha$ and $\theta$ can be made such that a robust
mirror-MOT is formed.

To realize a \ac{MOT} at a point $\pmb{r}$ with the crosswire geometry, the
quadrupole magnetic field must vanish at that point. We can apply an arbitrary
uniform bias field $\pmb{\beta}$ to cancel the field of the crosswires at
$\pmb{r}$, i.e., $\pmb{\beta} = -\pmb{B}(\pmb{r})$. By linearity, the residual
field is then
\begin{equation}
  \pmb{B} = \bigl(\hat{\pmb{G}}_1 + \hat{\pmb{G}}_2\bigr)\cdot\delta\pmb{r},
  \label{eq:BfromG}
\end{equation}
where $\hat{\pmb{G}}_{1,2}$ are the contributions from the two wires.

The magnetic field due to an infinitely long, thin wire passing through the
origin and carrying current $I$ along the unit vector $\hat{u}$ is
\begin{equation}
  \pmb{B}(\pmb{r}) =
  \frac{\mu_0 I}{2\pi}\,
  \frac{\hat{u}\times\pmb{r}_\perp}{r_\perp^{2}},
  \qquad
  \pmb{r}_\perp = \pmb{r} - (\pmb{r}\cdot\hat{u})\,\hat{u}.
  \label{eq:wirefield}
\end{equation}
Using this, it is straightforward to show that for our crosswire geometry
\begin{eqnarray}
  \hat{\pmb{G}}_1 +\hat{\pmb{G}}_2 &=&
  \frac{\mu_0 I_1}{2\pi }
  \begin{pmatrix}
    a_+ & -a_+ & b_+ \\
    -a_+ & a_+ & -b_+ \\
    b_+ & -b_+ & -2a_+
  \end{pmatrix} \nonumber \\ &+&
  \frac{\mu_0 I_2}{2\pi}
  \begin{pmatrix}
    a_- & a_- & b_- \\
    a_- & a_- & b_- \\
    b_- & b_- & -2a_-
  \end{pmatrix},
  \label{eq:G1G2}
\end{eqnarray}
where
$a_\pm = (r_z \pm T/2)(r_x\mp r_y)/\sqrt{2} / r_\pm^4$,
$b_\pm = ( (r_z \pm T/2)^2 - (r_x\mp r_y)^2/2 )/\sqrt{2}  / r_\pm^4$,
and
$r_\pm^2 = (x \mp y)^2/2 + (z \pm T/2)^2$.
In what follows, we hold the $z$-component $r_z$ fixed and treat $T$ as a
constant; thus $a_\pm$ and $b_\pm$ can be regarded as functions only of
$(r_x,r_y)$ (e.g., $a_\pm=a_\pm(r_x,r_y)$). To produce the desired \ac{MOT} fields, we
need to find positions $(r_x,r_y)$ and currents $(I_1,I_2)$ such that
Eq.~\eqref{eq:G1G2} matches Eq.~\eqref{eq:rotated}.

To obtain the zero entries in Eq.~\eqref{eq:rotated}
from the crosswire geometry, we require
\begin{equation}
  I_1 a_+ = I_2 a_-,\qquad I_1 b_+ = I_2 b_-,
  \label{eq:constraints}
\end{equation}
at the chosen \ac{MOT} position $\pmb{r}$. Under these conditions,
\begin{equation}
  \hat{\pmb{G}}
  = 2a_+\,\frac{\mu_0 I_1}{2\pi}
  \begin{pmatrix}
    1 & 0 & b_+/a_+ \\
    0 & 1 & 0 \\
    b_+/a_+ & 0 & -2
  \end{pmatrix}.
  \label{eq:fromcurr}
\end{equation}

Comparing Eqs.~\eqref{eq:rotated} and \eqref{eq:fromcurr}, the aspect parameter
must satisfy
\begin{equation}
  \alpha = 3\,\frac{1 - \cos 2\theta}{3 + \cos 2\theta}.
  \label{eq:alpha}
\end{equation}
Note that the aspect parameter is now constrained by the rotation angle.
Substituting Eq.~\eqref{eq:alpha} into Eq.~\eqref{eq:rotated} gives
\begin{equation}
  \hat{\pmb{G}} =
  \frac{4 \cos 2\theta}{3 + \cos 2 \theta}\, G
  \begin{pmatrix}
    1 & 0 & \tfrac{3}{2} \tan 2\theta \\
    0 & 1 & 0 \\
    \tfrac{3}{2} \tan 2 \theta & 0 & -2
  \end{pmatrix}.
  \label{eq:rotated2}
\end{equation}

Comparing Eq.~\eqref{eq:rotated2} and Eq.~\eqref{eq:fromcurr} yields a final
system of equations constraining the $a$, $b$, $c$, and $d$ parameters:
\begin{equation}
  \frac{b_+}{a_+} = \frac{3}{2} \tan 2\theta,\qquad \frac{a_+}{b_+} = \frac{a_-}{b_-}.
  \label{eq:conditions}
\end{equation}
We have verified numerically that experimentally realistic positions $(r_x,r_y)$
exist that solve the above system of equations for any angle $\theta$. Therefore,
the crosswire geometry can produce \ac{MOT} fields at any angle with the only
constraint being Eq.~\eqref{eq:alpha}, the relationship between the aspect
parameter and rotation angle.

\begin{figure}[h] \centering
\includegraphics[width=1\linewidth]{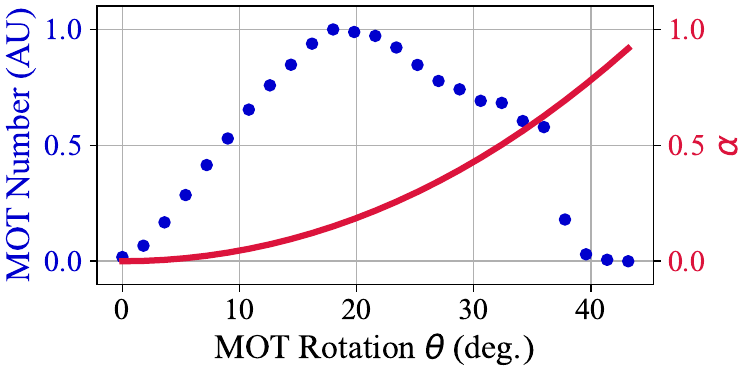}
    \caption{Aspect parameter and experimentally measured \ac{MOT} number as a function
    of rotation angle. As the rotation angle approaches $45^\circ$, the aspect
    parameter approaches 1, which is unsuitable for \ac{MOT} production. Due to
    this interdependence between angle and aspect parameter, the maximum \ac{MOT}
    number is found at an angle of $20^\circ$. } \label{fig:MOTNum_Alpha}
\end{figure}

Figure \ref{fig:MOTNum_Alpha} shows $\alpha$ as a function of $\theta$. An
aspect parameter of $\alpha = 0$ necessitates a $0^\circ$ rotation, which shows
that a crosswire can make an ideal traditional \ac{MOT} with fields aligned with the
chip axes. Meanwhile, a $45^\circ$ rotation leads to an aspect parameter of
$\alpha=1$ which is functionally a 2D \ac{MOT} instead of a 3D \ac{MOT}. Naively, one
might expect this result to prove that the crosswire geometry cannot be used for
a mirror-MOT. However, \acp{MOT} are robust, and a mirror-MOT does not require
exactly $45^\circ$ rotation or exactly $\alpha = 0$.

Now the question is: what is the optimal rotation angle/aspect parameter which
yields the largest \ac{MOT} number? We answered this question experimentally. Figure
\ref{fig:MOTNum_Alpha} shows the experimentally measured \ac{MOT} atom number as a
function of rotation angle. The magnetic fields and crosswire currents were
calculated for each angle by numerically solving Eq.~\eqref{eq:conditions} at a
fixed height. The optimal compromise between angle and aspect parameter was
found at $\theta = 20^\circ$, $\alpha = 0.185$.

\begin{figure} \centering
\includegraphics[width=1\linewidth]{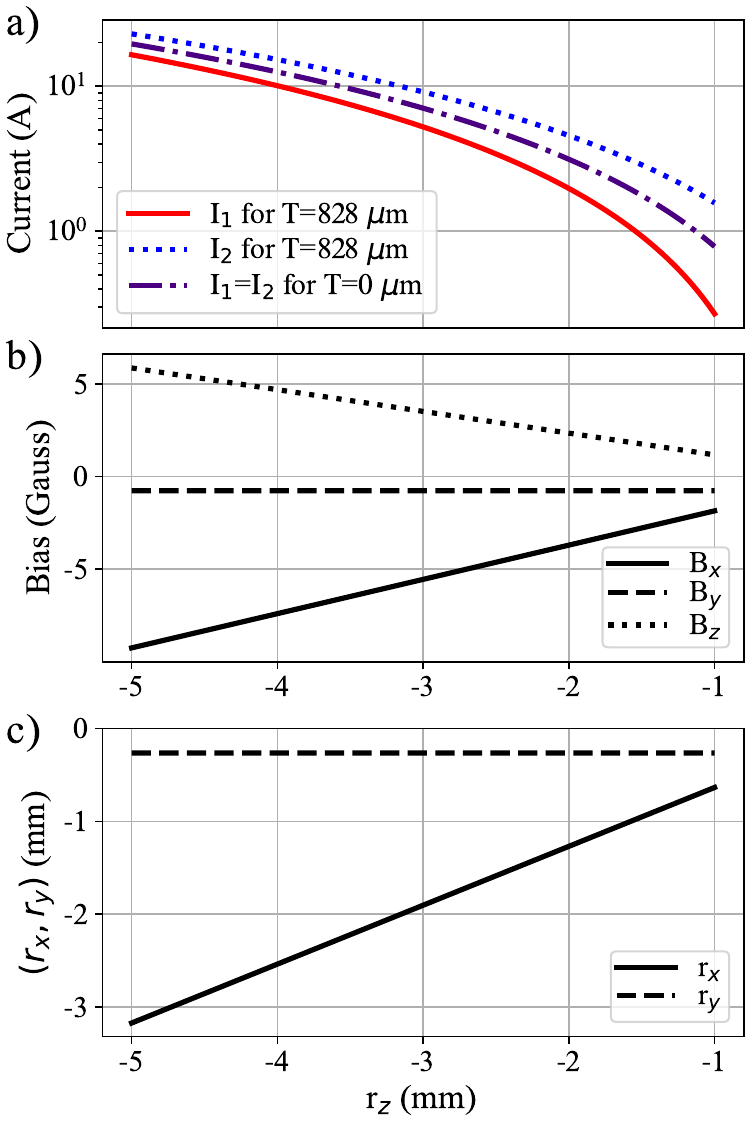}
    \caption{
    Currents and bias fields required to produce a \ac{MOT} at different
    distances from the chip ($r_z$). The angle for these calculations is set to
    $20^{\circ}$, but the results are similar for other angles. a) The currents
    needed in the crossed wires with the distance between wire centers ($T$) set
    to 828 $\mu$m and 0 $\mu$m. The actual experiment has a chip where $T=828$
    $\mu$m. The peak required currents range from about 20 A at a distance
    of -5 mm, to about 2 A at a distance of 1 mm. Such currents are quite
    feasible for a \ac{DBC} atom chip. b) Required bias fields are $\mathord{<} 6$ Gauss
    which is easy to achieve with Helmholtz coils. c) The $(r_x, r_y)$
    displacement needed to achieve a $20^{\circ}$ rotated gradient as a function
    of trap height. As the height decreases, so does the $(r_x, r_y)$
    displacement, which allows for easy mode-matching to a magnetic trap upon
    transfer closer to the atom chip.
    } \label{fig:Current_Bias} \end{figure}

Figure \ref{fig:Current_Bias} shows the currents and bias fields required to
produce a gradient with a $\theta = 20^\circ$ rotation as a function of trap
height. With reasonable currents, it is possible to produce a \ac{MOT} far from the
atom chip where a large capture volume is possible, as well as be close to the
atom chip where it is easier to mode match with the magnetic trap. The shape and
function of the magnetic trap produced by the crosswire is similar to the so-called dimple traps that are sometimes formed in the center of a z-wire
\cite{farkas2013efficient,horikoshi2006atom}. As the \ac{MOT} is brought closer to
the atom chip, the $(r_x,r_y)$ displacement also shrinks, as does the z-bias; this
is useful for mode matching since the magnetic trap is directly beneath the wire
cross with a z-bias $\approx 0$.

\section{Experimental Setup}

\begin{figure}[h!]
    \centering
    \includegraphics[width=.95\linewidth]{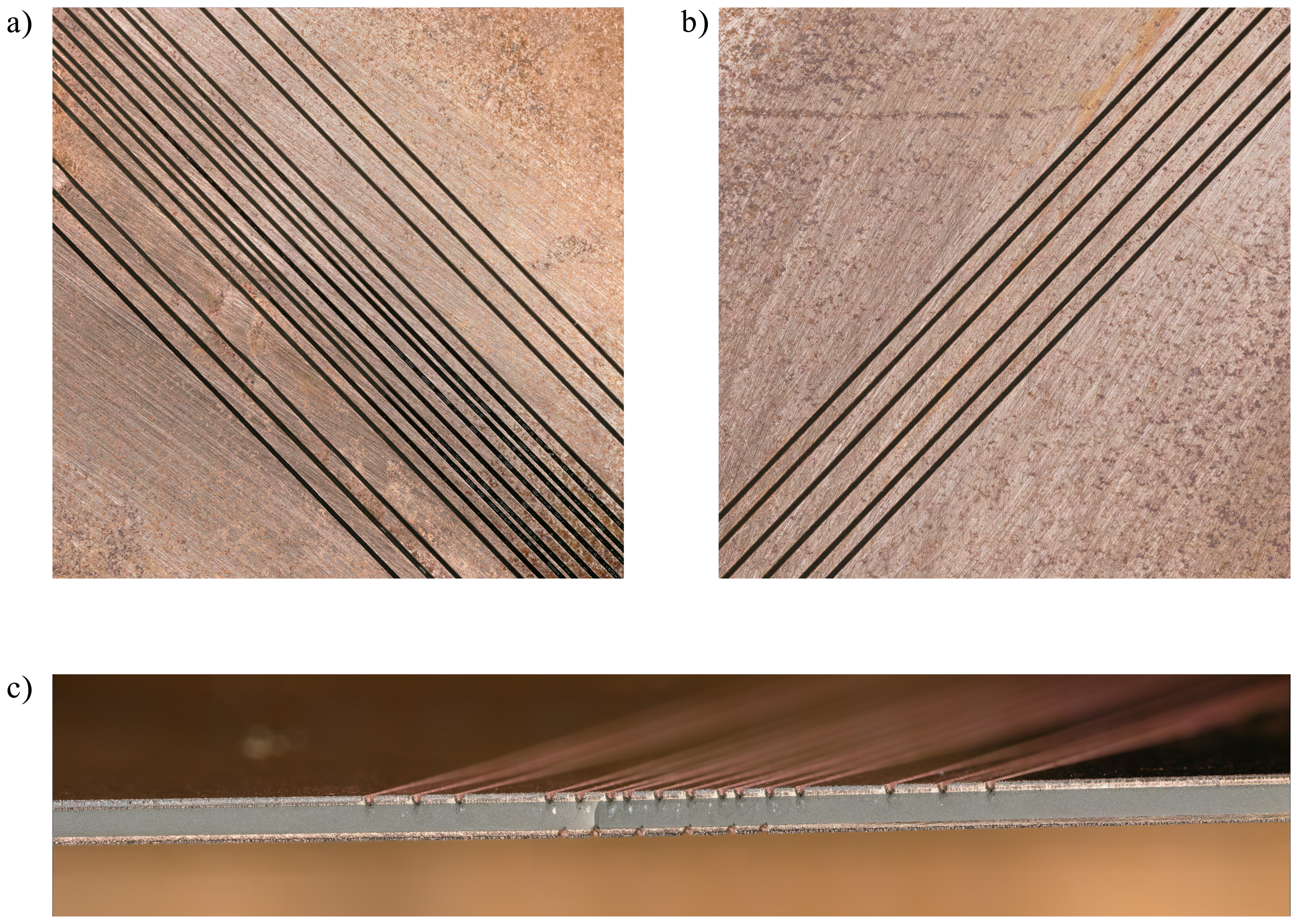} 
    \caption{
    Pictures of atom chip trace cuts. a) A top-down view of the atom chip
    showing the polynomial wire pattern. b) A top-down view of the atom chip
    showing the waveguide wire pattern c) A side view of a sectioned atom chip
    showing the relative size of the wires to the \ac{AlN} and copper layers, as
    well as the profile of the laser cut copper traces. The laser cut gap
    tapers as it moves from the edge of the \ac{Cu} down to the layer of
    \ac{AlN} at the bottom. The copper regions on the chip are considered
    isolated with the resistance between copper traces $\mathord{>}1\,$M$\Omega$.}
    \label{fig:Chips}
\end{figure}

Commercially available \ac{DBC} is used as the base substrate for the atom chips
\cite{squires2011_atomChipsDBC}. Although a range of thicknesses are available
(and viable), the \ac{DBC} used for these results has \ac{AlN} that is 625
$\mu$m thick, and \ac{Cu} layers that are each 203 $\mu$m thick. The atom chip
traces are created using laser etching to remove \ac{Cu} around the desired
current paths. We achieve high precision and repeatability, with wire widths as
small as $50\,\mu$m with a spacing between wires on the order of $150\,\mu$m.
Fig.~\ref{fig:Chips} shows top views and a sectioned view of a laser-etched atom
chip.

Since \ac{AlN} and \ac{Cu} both have thermal conductivity greater than 300
W/m$\cdot$K, the chips are able to disperse heat very effectively. This allows
us to run much higher currents ($>$50 A) as compared to other atom chips, which
typically operate at $\mathord{\sim}5$~A. The ability to run high currents allows
the chips to be placed outside of the vacuum chamber, which is useful for
iterating on chip design without breaking vacuum~\cite{squires2016_exVacuuo}.

To create a crossed-wire design, one trace is etched on the top of the chip, and
another trace is etched, perpendicular to the first trace, on the bottom of the
chip. This double-sided design allows current to be applied to the traces
independently. While the separate layers allow for electrical isolation, it adds
the requirement to apply different currents in the wires to produce the same
gradient at the height of the center of the quadrupole field.
Figure~\ref{fig:Current_Bias}a shows the difference in the top layer current,
which is higher, versus the lower layer current, as compared to the current that
would be required for a single-layer atom chip with a distance that is centered
in the atom chip. While we only discuss this crossed-wire arrangement, the
actual chip includes other wires for creating a polynomial waveguide
trap~\cite{stickney2017_tuneablePotentials} (see Figure~\ref{fig:Chips}).

After laser etching, the chips are connectorized via soldered wire leads. The
high thermal conductivity of the \ac{DBC} and the \ac{Cu} make soldering
difficult. The technique we have implemented to solder the leads is to tin the
ends of the wire leads, heat the chip on a hot plate to $\mathord{\approx}90\%$ of the
solder wetting temperature, and solder the tinned wire to the chip surface with a
soldering iron. After all the leads have been soldered to the chip, the wires are
potted to the chip with epoxy. We have had the most success with Aerothane 5753
A/B for potting.

\subsection{Benchtop Apparatus}\label{sec:Apparatus}

The basic structure of our experimental setup is described in
Ref.~\citenum{squires2016_exVacuuo}. In brief, a 2D+\ac{MOT} is created at one
end of a glass vacuum chamber. The flux of the 2D+\ac{MOT} exits through a
small aperture into an intermediate chamber with a series of baffles to improve
pumping of atoms that are not in the central low-velocity beam. The atomic beam
then enters the lower-pressure 3D \ac{MOT} chamber that has a thin silicon
membrane as the top wall of the vacuum chamber, with the atom chip resting just
above it (see Fig.~\ref{fig:SetupCartoon}). Outside of the chamber, three pairs
of coils in a Helmholtz configuration create uniform bias fields in three
orthogonal directions. The field from the center crossed wires on the atom chip,
combined with the bias fields and the lasers, create a 3D-\ac{MOT} or a magnetic
trap depending on the stage of the experiment.

\subsection{Trapping and Cooling Protocols}\label{sec:TrappingCooling}

\begin{figure}
    \centering
    \includegraphics[width=0.9\linewidth]{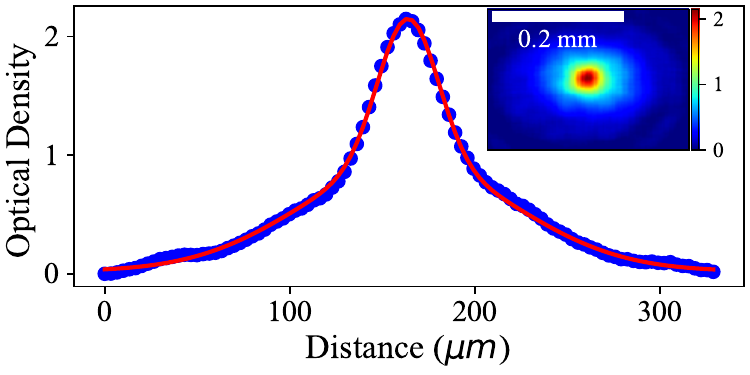} 
    \caption{
    Optical density of the atomic cloud after 10 ms time of flight. The main
    plot shows the density profile of an x-axis slice through the center of the
    cloud. The inset shows a picture of the full atomic cloud. A clear double-cloud structure is visible and evidence of partial condensation. From this
    cloud, we calculate a condensate fraction of 0.25. }
    \label{fig:Condensation}
\end{figure}

While the $20^{\circ}$ rotated \ac{MOT} gradient presented in
section~\ref{sec:Theory} serves as an effective starting point for producing a
crosswire-MOT, in practice, we optimize the \ac{MOT} number and loading into the
magnetic trap by feeding back on the number of atoms in the final magnetic trap.
After loading into the magnetic trap, we performed a quick optimization of forced
RF evaporation to demonstrate \ac{BEC} production in a crosswire magnetic trap loaded
by a crosswire-MOT. Figure \ref{fig:Condensation} shows a representative atom
cloud after 10~ms of time of flight with $4.5\times10^4$~atoms and a condensate
fraction of 0.25.

The typical experimental trapping parameters and protocols are: for loading the
3D-\ac{MOT}, we use 14.22~A in the top wire and 9.64~A in the bottom wire.
The bias fields $(B_x,B_y,B_z)$ are set respectively to (7.82, 1.63,
3.21)~Gauss. During \ac{MOT} loading, our cooling beam is detuned 8.2 MHz to the
red of the $F = 2 \rightarrow 3'$ transition. We also apply repump light,
resonant with the $F = 1 \rightarrow 2'$ transition.

After the \ac{MOT} loading stage, the \ac{MOT} is compressed and moved closer to
the chip by changing the currents and biases to 17.18~A in the top wire and
11.78~A in the bottom wire and bias fields (9.78, 1.03, -0.20)~Gauss. This is
followed by \ac{PGC}, and then the atoms are optically pumped into the
$|F=2;m_f=2\rangle$ magnetic sublevel.

After optical pumping, the atoms are transferred to an initial magnetic trap
($MT_{catch}$) with currents and biases set to 33.65~A in the top wire,
32.88~A in the bottom wire, and bias fields (28.03, 2.13, 1.62)~Gauss. This trap
is relatively weak with trap frequencies of (12,~82,~83)~Hz, but it has a large
enough capture area that it overlaps with the final position of the laser-cooled
atoms. After capture, the trap is compressed and brought closer to the chip
($MT_{comp}$) using 45~A in the top wire, 36~A in the bottom wire, and bias
fields (62, -2.5, -0.17)~Gauss, which results in trap frequencies of
(26,~138,~144)~Hz and a bottom field of 8.5~Gauss. The atoms are then cooled with
forced RF evaporation. As the atom cloud is cooled, the trap is moved closer to
the atom chip. The final trap at the end of evaporation is made using 40~A in
the top wire, and 25~A in the bottom wire, and bias fields (62, 0.3, -0.17)~Gauss,
which results in trap frequencies of (35, 418, 419)~Hz and a bottom field of
1.7~Gauss. The exact details of the changes to RF ramps and trap will not be
given, but the changes in trapping conditions largely follow the procedures
in \cite{chipBEC_Hansel2001, squiresDissertation2008}. After 3.5~s of cooling,
a \ac{BEC} is generated with $4.5\times10^4$~atoms with a condensate fraction of
up to 0.25. 

All steps in forming a \ac{BEC} are performed with the crossed wires with the
appropriate bias field. The magnetic field created by the crossed wires is able
to produce a quadrupole trap and then, with small modifications of the wire
currents and bias fields, is able to produce a magnetic trap with a non-zero trap
minimum (Ioffe traps). The significance of creating both types of traps with a
simple wire configuration is enabling to the manipulation and trapping of
ultra-cold atoms in a variety of scenarios.

\section{Conclusions}

In conclusion, a crossed pair of conductors with appropriate bias fields can be
used to create all of the stages typically used to create a \ac{BEC}. There is
no compromise in the atom number during the mirror \ac{MOT} phase, and the
ability to magnetically trap the atoms at a variety of heights while maintaining
the aspect ratio of the axes of the trap shows the utility and flexibility of
this atom chip configuration. In principle and with appropriate spacing, this
crossed-wire design can also be patterned to create an array of ultra-cold
gases.

\section*{Disclaimer}
The views expressed are those of the authors and do not
necessarily reflect the official policy or position of the Department of the Air
Force, the Department of War, or the US Government.

\bibliographystyle{apsrev4-2}
\bibliography{references}

\end{document}